\documentclass[epj]{svjour}
\usepackage{amsmath,epsfig}
\usepackage{graphicx}
\usepackage{dcolumn}
\usepackage{bm}
\usepackage{amssymb}
\usepackage{amsmath}%
\begin{document}
\title{The scenario of two families of compact stars}
\subtitle{1. Equations of state, mass-radius relations and binary systems}
\author{Alessandro Drago\inst{1}, Andrea Lavagno\inst{2} \and Giuseppe Pagliara\inst{1} \and Daniele Pigato\inst{2}}                     
\offprints{}          
\institute{Dip.~di Fisica e Scienze della Terra dell'Universit\`a di Ferrara
and INFN Sez.~di Ferrara, Via Saragat 1, I-44100 Ferrara, Italy \and  Department of Applied Science and Technology, Politecnico
di Torino and INFN, Sez.~di Torino, I-10129 Torino, Italy}
\date{Received: date / Revised version: date}
%
\abstract{We present several arguments which favor the scenario of two
  coexisting families of compact stars: hadronic stars and quark
  stars. Besides the well known hyperon puzzle of the physics of
  compact stars, a similar puzzle exists also when considering delta
  resonances. We show that these particles appear at densities close to twice saturation density
  and must be therefore included in the calculations of the hadronic
  equation of state. Such an early appearance is strictly related to
  the value of the L parameter of the symmetry energy that has
  been found, in recent phenomenological studies, to lie in the range $40<L<62$ MeV.
We discuss also the threshold for the formation
of deltas and hyperons for hot and lepton rich hadronic matter.
 Similarly to the
  case of hyperons, also delta resonances cause a softening of the
  equation of state which makes it difficult to obtain massive hadronic
  stars. Quark stars, on the other hand, can reach masses up to $2.75
  M_{\odot}$ as predicted by perturbative QCD calculations.  We then
  discuss the observational constraints on the masses and
  the radii of compact stars. The tension between the precise
  measurements of high masses and the indications of the existence of
  very compact stellar objects (with radii of the order of $10$ km) is relieved when assuming that very
  massive compact stars are quark stars and very compact stars are
  hadronic stars. Finally, we discuss recent interesting measurements
  of the eccentricities of the orbits of millisecond pulsars in low mass
  X-ray binaries. The high values of the eccentricities found in some
  cases could be explained by assuming that the hadronic star, initially
  present in the binary system, converts to a quark star due to the increase of its central density.
\PACS{
      {PACS-key}{describing text of that key}   \and
      {PACS-key}{di-scribing text of that key}
     } 
} 
\maketitle
\section{Introduction}
Ultra-relativistic heavy ions experiments have provided many
indications of the formation of a new phase of strongly interacting
matter, named quark gluon plasma, which is obtained by heating up
hadronic matter (with almost vanishing baryon density) to temperatures
of few hundreds MeV \cite{Adams:2005dq}.  In this state, the
fundamental degrees of freedom of QCD, quarks and gluons, are
deconfined. A very interesting question of nuclear and hadronic
physics concerns the possibility of the formation of a deconfined
phase also at large baryon densities and small temperatures. Natural
systems to look for this state of matter are neutron stars. In this
respect, the recent discoveries of two stellar objects
\cite{Demorest:2010bx,Antoniadis:2013pzd} with masses of
$M=2M_{\odot}$ are very promising: the larger the mass the larger the
baryon density in the core of the stars.  Those massive compact stars
are therefore the best systems for studying the structure of the QCD
phase diagram at high densities.

There is huge collection of theoretical and phenomenological
calculations aiming at establishing whether quark matter can form in these stellar objects
and which would be the possible associated observational signatures.
One can distinguish three possible scenarios:
i) there is only one family of compact stars which are hybrid stars
i.e. stars composed by hadronic matter at low density and quark matter at high density.
   See for instance Refs. \cite{Chen:2011my,Bonanno:2011ch,Zdunik:2012dj,Kurkela:2014vha} for recent calculations providing
   equations of state stiff enough to support a star of $2M_{\odot}$.
ii) high mass ``twin compact stars'' \cite{Benic:2014jia}. In this scenario most of the stars
    would be composed only of nucleonic matter. Stellar configurations with masses of about $2M_{\odot}$ could 
be composed either
  of nucleonic or of hybrid matter. The two coexisting stellar configurations have different radii:
  hybrid stars are more compact than their ``twin hadronic stars''.
iii) two separated families of compact stars \cite{Drago:2013fsa}: hadronic stars which can be very compact and
have a maximum mass of about $1.5-1.6 M_{\odot}$; quark stars which can be very massive, up to $2.75 M_{\odot}$, and have large radii.

The third scenario is the one we will present in this paper and in the
accompanying paper 2. We will discuss in particular the
phenomenological motivations in favor of this model and we will try to
analyze the predictions which distinguish this model from the other
two.  The paper is organized as follows: in Sec.2 we present the
calculations of the equations of state of hadronic and quark
matter. In Sec.3 we compare our theoretical results with the
observational constraints on the masses and the radii of compact
stars.  In Sec. 4 we discuss the information one can obtain on the
Equation of State (EoS) from the study of compact stars in binary systems.
Final discussions and conclusions are presented in Sec. 5.

\section{Equations of state}
In this Section we present the EoS for hadronic
and quark matter used in the present investigation. 
For hadronic matter a problem hugely discussed in the literature
concerns how to reconcile the unavoidable appearance
of hyperons (at densities of the order of $2-3 n_0$) 
with the existence of compact stars with masses of $2 M_{\odot}$.
This problem is discussed in the contributions of Chatterjee and Vidana and of Oertel et al. in this volume. 
The similar problem of the appearance of delta resonances at finite
density is much less discussed in the literature for reasons 
that we will clarify in the following.
Here we summarize our findings concerning the simultaneous
formation of hyperons and deltas within a relativistic mean field
approach. 

In the first
subsection, the $\beta$-stable hadronic EoS is studied in the regime of
zero temperature within a nonlinear relativistic Walecka-type model
and in the framework of the so-called SFHo parametrization which takes
into account the recent experimental constraints. In the second
subsection, we extend the study of the SFHo model at finite value of
entropy per baryon.

\subsection{Hadronic equations of state including $\Delta$-isobars and hyperons at $T=0$}

Concerning the hadronic EoS we consider in this paper
two different relativistic EoS with the inclusion of the octet of
lightest baryons (nucleons and hyperons) and $\Delta$(1232) isobar
resonances. First, we study the nonlinear GM3 model of
Glendenning-Moszkowsky in which the interaction between baryons is
mediated by the exchange of a scalar meson $\sigma$, an isoscalar
vector meson $\omega$ and a isovector vector $\rho$
\cite{Glendenning:1991es}. Let us note that, within the GM3
parametrization, only the experimental value of the symmetry energy at
saturation $S$ is used to fix the coupling between the $\rho$ meson
and the nucleons. However, recently, a remarkable concordance among
experimental, theoretical, and observational studies has been found
\cite{Lattimer:2012xj}, by allowing to significantly constrain also
the value of $L$, the derivative with respect to the density of the
symmetry energy $S$ at saturation:
\begin{equation}
L=3 n_B
\frac{\mathrm{d}S}{\mathrm{d}n_B}\mid_{n_B=n_0}\, .
\end{equation}
Therefore, extensions of the GM relativistic mean-field model have
been implemented which include $\rho$ meson self-interaction
terms. These new parametrization modify the density dependence of the
symmetry energy at supranuclear densities and satisfy all of the
experimental constraints both from terrestrial and astrophysical data
by restricting $L$ to the range of 40 MeV $\lesssim L \lesssim$ 62 MeV
\cite{Lattimer:2012xj} \footnote{Notice that this rather restricted range for the values of $L$
is still debated, see for instance Ref.\cite{Tsang:2012se} where a looser constraint is proposed.}. 
To this purpose, we are going to compare the
results in the framework of the GM3 model (with a value of $L\simeq$
80 MeV, automatically fixed once a specific value of $S$ is adopted)
with a more sophisticated EoS, called SFHo, for which $S$ = 32 MeV
(very close to the GM3 value) and $L$ = 47 MeV
\cite{Steiner:2004fi,Steiner:2012rk}.

In the GM3 model the general form of lagrangian is given by \cite{Glendenning:1991es}
\begin{eqnarray}
{\cal L}_{\mathrm{octet}} &=& \sum_k \bar\Psi_{k}\left(i\gamma_\mu{\partial^\mu} - m_k
+ g_{\sigma k} \sigma - g_{\omega k} \gamma_\mu \omega^\mu
- \right. \nonumber \\ && \left. g_{\rho k}
\gamma_\mu\frac{{\mbox{\boldmath $\tau$}}_k}{2} \cdot
{\mbox{\boldmath $\rho$}}^\mu
\right)\Psi_k
+ \frac{1}{2}\left( \partial_\mu \sigma\partial^\mu \sigma
- m_\sigma^2 \sigma^2\right)\nonumber \\ &&
 -\frac{1}{4} \omega_{\mu\nu}\omega^{\mu\nu}
+\frac{1}{2}m_\omega^2 \omega_\mu \omega^\mu
- \frac{1}{4}{\mbox {\boldmath $\rho$}}_{\mu\nu} \cdot
{\mbox {\boldmath $\rho$}}^{\mu\nu} \nonumber \\
&&+ \frac{1}{2}m_\rho^2 {\mbox {\boldmath $\rho$}}_\mu \cdot
{\mbox {\boldmath $\rho$}}^\mu  + U(\sigma,\omega,{\mbox {\boldmath $\rho$}})\, ,
\label{lagm}
\end{eqnarray}
where the index $k$ runs over the baryon octet, $m_k$ is the bare mass
of the baryon $k$, ${\mbox{\boldmath $\tau_{k}$}}$ is the isospin
operator and finally $U$ is the mesons potential which can contain non
linear interaction terms.

Concerning hyperons, with the exception of the $\Lambda$, their binding energies in hypernuclei are highly uncertain (see, for example, Ref. \cite{Raduta:2014lja} and references therein) and thus also their couplings with mesons are poorly constrained.  Here, we use the parameters set of Refs.~\cite{Schaffner:1993nn,Schaffner:1995th,Drago:2013fsa} obtained by reproducing the following values of the binding energies in nuclear matter $U^N_i$:
\begin{equation}
U^N_{\Lambda}=-28\,\mathrm{MeV}, \, U^N_{\Sigma}=30\,\mathrm{MeV}, \, U^N_{\Xi}=-18\,\mathrm{MeV.}
\end{equation}
For the coupling with vector mesons we use the SU(6) symmetry relations:
\begin{eqnarray}
\frac{1}{3}g_{\omega N}=\frac{1}{2}g_{\omega \Lambda}&=&\frac{1}{2}g_{\omega \Sigma}=g_{\omega \Xi}\\
g_{\rho N}=\frac{1}{2}g_{\rho \Sigma}&=&g_{\omega \Xi},\,\,\,g_{\rho \Lambda}=0 \, .
\end{eqnarray}

In relativistic heavy ion collisions, where large values of temperature and density can be reached, a state of resonance matter may be formed and the $\Delta(1232)$-isobars are expected to play a central role. \cite{Zabrodin:2009fz,Hofmann:1994gn,Bass:1998vz,Lavagno:2010ah,Lavagno:2012bn}.
Moreover, it has been pointed out that the existence of $\Delta$s can be very relevant also in the core of neutron stars
\cite{Huber:1997mg,Xiang:2003qz,Chen:2007kxa,Chen:2009am,Schurhoff:2010ph}.

The mean-field Lagrangian density for the $\Delta$-isobars can be then expressed as
\begin{eqnarray}
{\mathcal L}_\Delta=\overline{\psi}_{\Delta\,\nu}\, \left[ i\gamma_\mu
\partial^\mu -(m_\Delta-g_{\sigma\Delta}
\sigma)-g_{\omega\Delta}\gamma_\mu\omega^\mu  \right. \nonumber \\ \left. -g_{\rho \Delta}
\gamma_\mu I_{3} \rho^{\mu}_3 \right] \psi_{\Delta}^{\,\nu} \, ,
\end{eqnarray}
where $\psi_\Delta^\nu$ is the Rarita-Schwinger spinor for the
$\Delta$-isobars ($\Delta^{++}$, $\Delta^{+}$, $\Delta^0$,
$\Delta^-$) and $I_{3}=\mathrm{diag}(3/2,1/2,-1/2,-3/2)$ is the matrix containing the isospin charges of the $\Delta$s.

As customary, for the couplings of hyperons and $\Delta$ isobars with
the mesons, we introduce the ratios 
\begin{equation}
x_{\sigma i}=g_{\sigma\,i}/{g_{\sigma N}}\,, \ \ \ x_{\omega i}=g_{\omega\, i}/{g_{\omega N}} \ \ \ x_{\rho i}=g_{\rho\, i}/{g_{\rho N}}\, ,
\end{equation}
where the index $i$ runs over
all the hyperons and $\Delta$ isobars.

Concerning the values of the $\Delta$-meson couplings, if the SU(6) symmetry is exact, one adopts the universal couplings $x_{\sigma\Delta}=x_{\omega\Delta}=1$.
As already extensively discussed in Ref. \cite{Glendenning:1984jr},
among the four $\Delta$ isobars, the $\Delta^-$ is likely to appear
first because it can replace a neutron and an electron at the top of their Fermi seas in $\beta$-stable matter. However, 
this particle is ``isospin unfavored'' because its isospin charge $t_3=-3/2$ has the
same sign of the isospin charge of the neutron. For large values of
the symmetry energy $S$ and, therefore, of $g_{\rho \Delta}$, the
$\Delta^-$ appears at very large densities or it does not appear at
all in dense matter thus playing no role in compact stars. Indeed, in
Ref. \cite{Glendenning:1984jr} the $\Delta$-isobars could appear in
neutron stars only for not physical small values of the symmetry
energy, obtained by setting $g_{\rho i}=0$ for all the baryons.
However, as already observed, in the GM3 model the
coupling $g_{\rho N}$ is fixed by using the experimental value of the symmetry energy, the most
recent estimates ranging in the interval $29 \lesssim S \lesssim 32.7$
MeV \cite{Lattimer:2012xj}. In this scheme no experimental information
on the density dependence of the symmetry energy can be incorporated
and in particular the $L$ parameter is automatically fixed once a
specific value of $S$ is adopted. It turns out that in the models
introduced in Refs. \cite{Glendenning:1991es,Glendenning:1984jr}, $L
\sim 80$ MeV and it is thus significantly higher than the values suggested by the most
recent analysis \cite{Lattimer:2012xj}.

Moreover, in this context let us observe that the SU(6) symmetry is
not exactly fulfilled and one may assume the scalar coupling ratio
$x_{\sigma\Delta}>1$ with a value close to the mass ratio of the
$\Delta$ and the nucleon \cite{Kosov:1998gp}. On the other hand, QCD
finite-density sum rule results show that the Lorentz vector
self-energy for the $\Delta$ is significantly smaller than the nucleon
vector self-energy implying therefore $x_{\omega\Delta}<1$
\cite{Jin:1994vw}.

In the many body analysis of Ref.\cite{Oset:1987re}, the real part of
the $\Delta$ self-energy has been evaluated to be about $-30$ MeV at
$n_B=0.75\, n_0$. Notice that this self energy is relative to the one
of the nucleon and the total potential felt by the $\Delta$ is the sum
of its self energy and of the nucleon potential, a number of the order
of $-80$ MeV. Also phenomenological analysis have been performed of
data from electron-nucleus
\cite{Koch:1985qz,Wehrberger:1989cd,O'Connell:1990zg}, photo-absorption
\cite{Alberico:1994sx} and pion-nucleus scattering
\cite{Horikawa:1980cv,Nakamura:2009iq}. Such analyses suggest a more
attractive interaction of the $\Delta$ in the nuclear medium with
respect to the nucleon one (see Refs.
\cite{Drago:2014oja,PhysRevC.92.015802} for more details). New
analyses, and possibly new experiments, aiming at a better
determinations of these couplings would be extremely important.
Notice also that no information is available for $x_{\rho\Delta}$
which in principle could be extracted by analyzing scattering on
neutron rich nuclei (see the recent discussion in \cite{Li:2015hfa,Benlliure:2015qea}).

The threshold for the formation of the i-th baryon is given
by the following relation:
\begin{equation}
\mu_i \ge m_i-g_{\sigma i} \sigma + g_{\omega i}\omega + t_{3 i}g_{\rho i} \rho \, ,
\label{soglia}
\end{equation}
where $\sigma$, $\omega$ and $\rho$ are the expectation values of the
corresponding fields, $\mu_i$, $m_i$ and
$t_{3 i}$ are the chemical potential, the mass and the isospin charge
of the baryons.  The baryon chemical potential $\mu_i$ are obtained by
the $\beta$-equilibrium conditions: 
\begin{equation}
\mu_i=\mu_B+c_i\,\mu_{C}\, ,
\end{equation}
where $\mu_B$ and $\mu_{C}$ are the chemical potentials associated with the
conservation of the baryon number and the electric charge respectively
and $c_i$ is the electric charge of the i-th baryon.

In Fig.~1, we display the baryon density dependence of the particle's
fractions in the GM3 model for $x_{\sigma\Delta}=1.25$,
$x_{\omega\Delta}=1$ and neglecting in this case the coupling of the
$\Delta$-isobars with the $\rho$ meson ($x_{\rho\Delta}=0$).  Let us
note that in this scheme the appearance of the $\Delta$-isobars is a
consequence of the introduction of a more attractive interaction
($x_{\sigma\Delta}>1$) of $\Delta$-particles with respect to the
nucleon in the mean field approximation, as in Refs.
\cite{Kosov:1998gp,Jin:1994vw,Li:1997yh}.  It is remarkable that the
early appearance of $\Delta$ resonances, the first one being the
$\Delta^-$, considerably shifts the onset of hyperons which start to
form at densities of $\sim 5 \,\rho_0$ (see the curve for the
$\Lambda$'s).

\begin{figure}[ptb]
\vskip 0.5cm
\begin{centering}
\epsfig{file=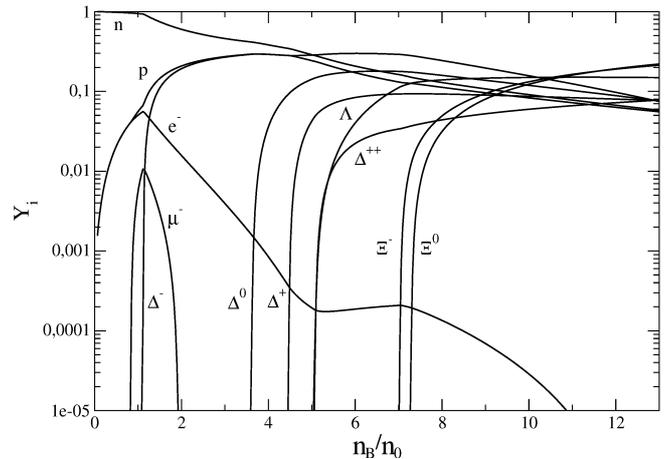,height=6cm,width=8.5cm,angle=0}
\caption{Particles fractions as functions of baryon density (in units of the nuclear saturation density $n_0$) in the GM3 model for $x_{\sigma\Delta}=1.25$, $x_{\omega\Delta}=1$.}
\end{centering}
\end{figure}

With the purpose of including 
the new experimental constraints about the value of the density derivative of the symmetry energy $L$, we can first study an extended GM3 model by considering the following density dependent baryon-$\rho$ meson coupling \cite{Drago:2014oja,Typel:2009sy}
\begin{equation}
g_{\rho i}=g_{\rho  i}(n_0)e^{-a(n_B/n_0-1)} \, .
\end{equation}
In this way we introduce a single parameter $a$ which affects only the value of $L$ leaving untouched the other properties of nuclear matter at saturation.

\begin{figure}[htb]
\vskip 0.5cm
\begin{centering}
\epsfig{file=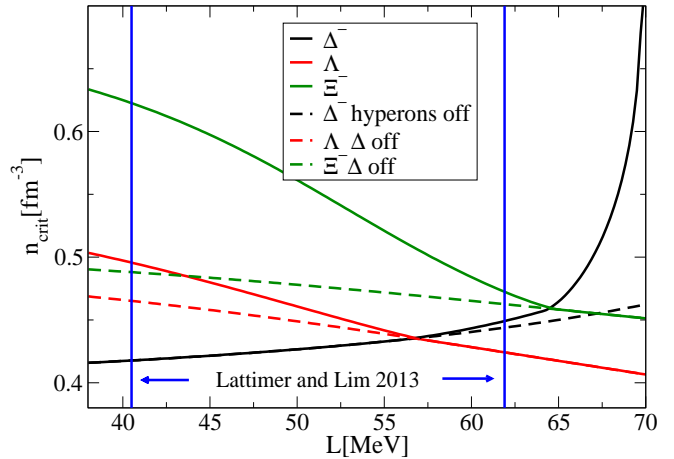,height=8.5cm,width=6cm,angle=-90}
\caption{Threshold densities of hyperons and $\Delta$s as functions of the $L$ parameter
for $x_{\sigma\Delta}=x_{\omega\Delta}=x_{\rho\Delta}=1$ within the modified GM3 model with the introduction of a density dependent baryon-$\rho$ meson coupling (see the text for details). The continuous lines
refer to the case in which all the degrees of freedom are included in the computation of the EoS and the dashed lines refer to the case in which either hyperons or $\Delta$s are artificially switched off. The vertical lines
indicate the range of allowed values of $L$, as found in \cite{Lattimer:2012xj}.}
\label{soglie}
\end{centering}
\end{figure}

The role of such density dependent baryon-$\rho$ meson coupling in the
modified GM3 model can be observed in Fig. \ref{soglie}, where is
reported the value of $n_{\rm crit}^B$ for the different baryons
as a function of $L$.  We limit this first discussion to the case of the
$\Lambda$, $\Delta^-$ and $\Xi^-$ which are the first heavy baryons
appearing as the density increases (notice that $\Sigma$ hyperons are
unfavored due to their repulsive potential) in the so-called universal
coupling $x_{\sigma\Delta}=x_{\omega\Delta}=x_{\rho\Delta}=1$. One can
notice the different behavior of the thresholds: the larger the value
of $L$ the larger $n_{\rm crit}^{\Delta}$ and the smaller $n_{\rm
  crit}^{\Lambda}$ and $n_{\rm crit}^{\Xi^-}$.

At high values of $L$, larger than about $65$ MeV, the threshold of
the $\Delta^-$ increases very rapidly with $L$. This corresponds to
the values of $L$ for which the $\Xi^-$ appears before the $\Delta^-$
thus completely suppressing those particles. Indeed within the GM3
model, for which $L\sim 80$ MeV, the $\Delta^-$ do not appear at all
as already found in Ref. \cite{Glendenning:1984jr}. Similarly, one can
notice that if the isobars are formed before the hyperons, what
happens below $L\sim 56$ MeV, $n_{\rm crit}^{\Lambda}$ and $n_{\rm
  crit}^{\Xi^-}$ are shifted to larger densities, as already noticed
in Ref. \cite{Drago:2013fsa}. Analogues results have been found in
Ref. \cite{Glendenning:1984jr}, where two cases are analyzed,
corresponding to a finite and to a vanishing value of $g_{\rho N}$,
with the result that in the case of $g_{\rho N}=0$ the isobars are
favored. The blue lines mark the range of the values of $L$ indicated
by the analysis of Ref.~\cite{Lattimer:2012xj}. Therefore, the recent
constraints on $L$ imply that at densities close to three times $n_0$
both the hyperons and the isobars must be included in the equation of
state and for the lower allowed values of $L$, the isobars appear even
before the hyperons. Finally, let us stress that in this analysis we
have chosen a rather conservative choice for the couplings between
$\Delta$s and mesons. If higher values of $x_{\sigma\Delta}$ and or
lower values for $x_{\omega\Delta}$ are adopted, $n_{\rm
  crit}^{\Delta}$ can result to be smaller than $n_{\rm
  crit}^{\Lambda}$ and $n_{\rm crit}^{\Xi^-}$ for all the acceptable
values of $L$.

In comparison with the previous results, we consider now a more
sophisticated model for the EoS proposed in
Ref. \cite{Steiner:2004fi,Steiner:2012rk}, where we use the
parametrization called SFHo for which $S=32$ MeV (very close to the
GM3 value) and $L=47$ MeV with the addition of hyperons (assuming
SU(6) symmetry) and $\Delta$-isobar degrees of freedom (assuming
$x_{\sigma\Delta}=x_{\rho\Delta}=1$ and different values for
$x_{\omega\Delta}$).

\begin{figure}[htb]
\vskip 0.5cm
\begin{centering}
\epsfig{file=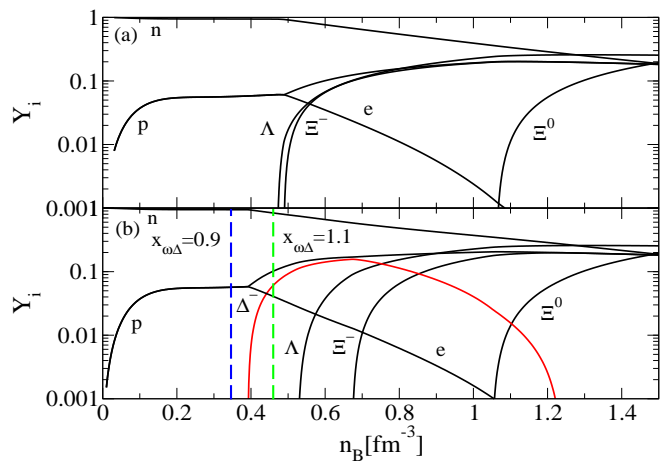,height=8.5cm,width=6cm,angle=-90}
\caption{Particles fractions as functions of the baryon density within the SFHo model: only hyperons (panel (a)),
hyperons and $\Delta$s (panel (b)) for $x_{\sigma\Delta}=x_{\omega\Delta}=x_{\rho\Delta}=1$. The red line indicates
the fraction of the $\Delta^-$ which among the four $\Delta$s are the first to appear.
The blue and the green vertical lines indicate the onset of the formations of $\Delta^-$ for $x_{\omega\Delta}=0.9$ and  $x_{\omega\Delta}=1.1$, respectively.}
\label{particles}
\end{centering}
\end{figure}

Results for the particles' fractions as function of the baryon density
in $\beta$-stable matter within the SFHo model are displayed in
Fig. \ref{particles}. In the upper panel, we have included only
hyperons: the $\Lambda$ and the $\Xi^-$ appear at a density of about
$0.5$ fm$^{-3}$ and then the $\Xi^0$ at a density of about $1.1$
fm$^{-3}$. In the lower panel we include also the $\Delta$ isobars. In
agreement with the findings of the previous analysis, for small values
of $L$ the $\Delta$s appear at densities relevant for neutron stars
and actually, in the SFHo model, they appear even before the hyperons
with the $\Delta^-$ formed at a density of about $0.4$ fm$^{-3}$. The
appearance of these particles delays the appearance of hyperons.  It
is important to remark that, within the SFHo model, even using more
repulsive interaction than nucleons, $x_{\omega\Delta}=1.1$, the
$\Delta^-$ appear before hyperons.

In conclusion, the early appearance of $\Delta$-isobars results to be
strictly related to the value of the $L$ parameter of the symmetry
energy and, for the values of $L$ indicated in Ref.\cite{Lattimer:2012xj}, such particle degrees of
freedom influence the appearance of hyperons and cannot be neglected
in the EoS. These results have been confirmed in the more recent Ref.\cite{Cai:2015hya}.

\subsection{Hadronic equation of state at finite entropy per baryon}

In this subsection we are going to study the behavior of the hadronic
EoS for conditions realized in protoneutron stars (PNS).
In particular, we focus our
investigation by considering the more realistic SFHo parametrization
in the first stage of the protoneutron stars (PNS) evolution,
corresponding to a total entropy per baryon equal to one, in which
neutrinos are trapped and strongly influence the chemical composition
of the PNS. Therefore, we also take into account of leptons particle
by fixing the lepton fraction
\begin{equation}
Y_{L}=Y_e+Y_{\nu_e}=(n_{e}+n_{\nu_e})/n_{B} \, ,
\end{equation}
where $n_e$, $n_{\nu_e}$ and $n_B$ are the electron, neutrino and
baryon number densities, respectively. 

The total entropy per baryon is calculated by means of 
\begin{equation}
s=\frac{S_B+S_l}{T\, \rho_B} \, ,
\end{equation}
where $S_B=P_B+\epsilon_B -
\sum_{i=B}\mu_i\rho_i$ and $S_l=P_l+\epsilon_l-
\sum_{i=l}\mu_i\rho_i$, and the sums are extended over all the
baryons and leptons species.

\begin{figure}[htb]
\begin{center}
\resizebox{0.48
\textwidth}{!}{%
\includegraphics{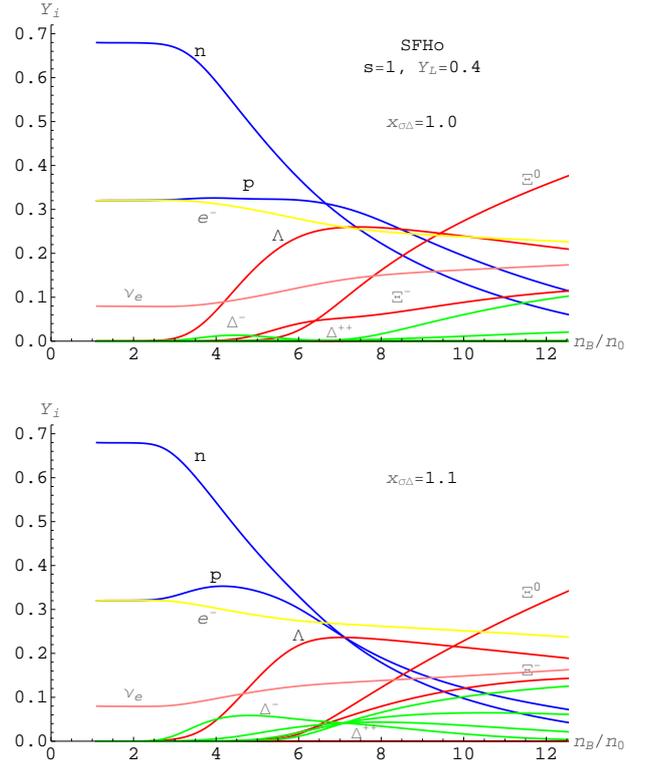}
}
\vskip 0.4cm
\caption{Particle concentrations $Y_i$ as a function of the baryon density for
$s=1$ and $Y_{L}=0.4$ in the SFHo parametrization with the coupling $x_{\sigma\Delta}=1.0$ (upper panel) and $x_{\sigma\Delta}=1.1$ (lower panel). In both cases $x_{\omega\Delta}=x_{\rho\Delta}=1$.}
\label{fig_yrhob}
\end{center}
\end{figure}

It is well known that the presence of trapped neutrinos significantly
alter the protons and the electrons abundance and strongly influence
the threshold of hyperons formation. This is also true in the presence
of $\Delta$-isobar degrees of freedom.  With the purpose of investigating
this problem, in Fig. \ref{fig_yrhob}, we report the particle
concentrations $Y_i$ as a function of the baryon density for $s=1$ and
$Y_{L}=0.4$ in the SFHo parametrization with the coupling
$x_{\sigma\Delta}=1.0$ (upper panel) and $x_{\sigma\Delta}=1.1$ (lower
panel).

Let us observe that, in the case of the universal coupling
$x_{\sigma\Delta}=x_{\omega\Delta}=x_{\rho\Delta}=1$ (upper panel),
$\Lambda$ and $\Delta^-$ particles appear approximately at the same
baryon density ($n_B\approx 3 \, n_0$). For $x_{\sigma\Delta}=1.1$
(lower panel), the onset of $\Delta^-$ particles is shifted at lower
densities and the presence of $\Delta$-isobars become more
relevant. In both cases, the population of strange $\Lambda$ particle
becomes relevant (greater than $5\%$) at about $n_B\approx 4 \,n_0$.

\begin{figure}[htb]
\begin{center}
\vskip 0.4cm
\resizebox{0.48
\textwidth}{!}{%
\includegraphics{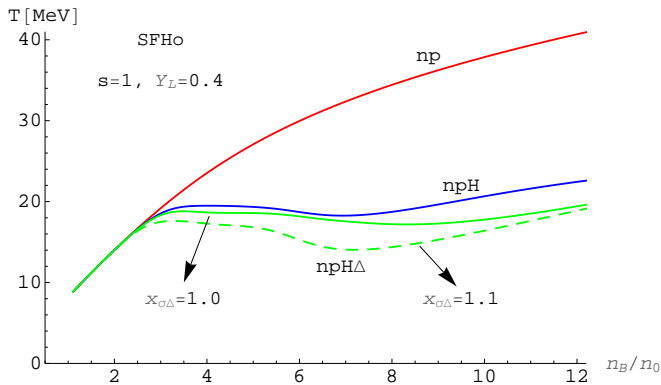}
} \caption{Temperature as a function of the baryon density (in units of $n_0$) at fixed entropy per baryon
and lepton fraction ($s=1, Y_{L}=0.4$) for the SFHo parametrization. The labels $np$, $npH$ and $npH\Delta$ stand for nucleons, nucleons plus hyperons, nucleons plus hyperons and $\Delta$-isobars, respectively. The results for two different meson-$\Delta$ couplings in the curves $npH\Delta$ are reported, the continuous line refers to $x_{\sigma\Delta}=1.0$ while the dashed line stands for $x_{\sigma\Delta}=1.1$.}
\label{temp_rhob}
\end{center}
\end{figure}

The features observed in the particle concentration are also reflected
in Fig. \ref{temp_rhob}, where we show the temperature as a function
of the baryon density for nucleonic matter ($np$), for hyperonic
matter ($npH$) and with the inclusion of $\Delta$-isobar degrees of
freedom ($npH\Delta$). As before, in the presence of $\Delta$
particles, we have considered two different meson-$\Delta$ couplings
(continuous line for $x_{\sigma\Delta}=1.0$ and dashed line for
$x_{\sigma\Delta}=1.1$). For $x_{\sigma\Delta}=1.0$, $\Lambda$ and
$\Delta^-$ particles start at $n_B\approx 3 \, n_0$ and, for a large
baryon density range, the behavior is almost isothermal ($T\approx
18\div 20$ MeV). A more discontinuous behavior can be observed in the
case of $x_{\sigma\Delta}=1.1$, due to the presence of the four
$\Delta$-isobar states.

\begin{figure}[htb]
\begin{center}
\resizebox{0.49
\textwidth}{!}{%
\includegraphics{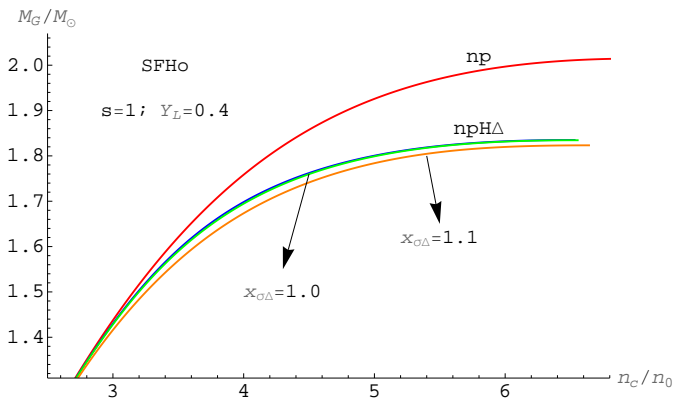}
}
\vskip 0.8cm
\caption{Gravitational mass as a function of the central baryon density $n_c$ for
$s=1$ and $Y_{L}=0.4$ in the SFHo parametrization with the couplings $x_{\sigma\Delta}=1.0$ and $x_{\sigma\Delta}=1.1$.}
\label{fig_mg_rhoc}
\end{center}
\end{figure}

In Fig. \ref{fig_mg_rhoc}, the gravitational mass as a function of the
central baryon density $n_c$ for $s=1$ and $Y_{L}=0.4$ is reported for
two different values of the $x_{\sigma\Delta}$ coupling ratio. In this
case no appreciable difference can be observed for $npH$ and
$npH\Delta$ curves (overlapped blue and green curves in the figure),
except for a greater central density $n_c$ reached in presence of
$\Delta$ particles. In agreement with the previous results, we can see
from the figure that hyperons and $\Delta$s appear at $n_B\approx 3 \,
n_0$, corresponding to $M_G\approx 1.45 M_\odot$. 
For stellar configurations with masses below this value,
deltas and hyperons do not appear or play a marginal role.
Therefore we do not expect any difference concerning the SN explosion
mechanism in the two families scenario with respect to the standard one.
For larger masses there are a few possibilities:
\begin{itemize}
\item hyperons appear and trigger the transition to quark matter halting the collapse;
\item hyperons appear but their abundance is not large enough to trigger the conversion
and a black hole will form after deleptonization \cite{Prakash:1996xs}.
\end{itemize}
In this little discussion we have not taken into account 
the rotation of the progenitor what can play an important
role as discussed in paper 2.

\subsection{Quark matter equation of state}
The quark matter EoS at densities reachable in the core
of compact stars is basically completely unknown. In the literature,
bag models or chiral models at finite chemical potential have been
widely used which capture two important non-perturbative aspects of
QCD: confinement the former and chiral symmetry breaking the latter
\cite{Buballa:2003qv}. Alternatively, one can resort to perturbative
QCD calculations: it has been shown that at high temperature,
vanishing chemical potential and finite quark masses perturbative
calculations provide results consistent with lattice QCD for
temperature larger than about $0.2$ GeV
\cite{Kurkela:2009gj,Fraga:2013qra}. The same technique has been used
to compute the EoS at finite chemical potential and
vanishing temperature with the very interesting result that the
strange quark matter EoS could be stiff enough to
support stars with $2.75 M_{\odot}$ \cite{Kurkela:2009gj}. Here, we
adopt the parametrization of the pQCD calculations at finite chemical
potential presented in \cite{Fraga:2013qra}. This parametrization has
only one free parameter, the scale parameter $X$, which is the ratio
between the renormalization scale and the baryon chemical potential
and it ranges between 1 and 4.  We choose here the value $X=3.5$ for
which the maximum mass of quark stars is $2.53 M_{\odot}$ (see
solid red line in Fig.\ref{massa-raggio}).
It is important to remark that all the calculations of the quark
matter EoS, such as the ones of \cite{Kurkela:2009gj}, are affected by
large uncertainties which inevitably affect also the predicted value
of the maximum mass. For instance, also in the simple MIT bag model
parametrization of Ref.\cite{Weissenborn:2011qu}, it has been proven
that very massive configurations can be obtained when varying the free
parameters of the model.  Moreover, it has been shown that, color
superconductivity, which is a purely non-perturbative phenomenon,
helps in obtaining high masses. 
Recently, within a SU(3) quark meson model it has also been shown that
massive quark stars configurations are allowed although in a specific
parameters region \cite{Zacchi:2015lwa}. Clearly, new and precise
astrophysical measurements are needed to constrain the properties of
dense matter.

\section{Masses and radii: theory vs observations}
Let us discuss the information on the mass-radius relation obtained by
means of astrophysical observations.  The direct and precise
measurements of the masses of PSR J1614-2230 with $M=1.97\pm
0.04M_{\odot}$ \cite{Demorest:2010bx} and of PSR J0348+0432, with
$M=2.01\pm 0.04M_{\odot}$ \cite{Antoniadis:2013pzd}, clearly represent
the most important constraints that theoretical calculations must
fulfill. A possible candidate with a mass even larger already exists:
it is the black widow pulsar PSR B1957+20 whose mass could be of
$2.4\pm 0.12 M_{\odot}$ provided that the modeling of the light curves
is correct \cite{vanKerkwijk:2010mt}. Taking into account the
systematic uncertainties on the light curves fit, the lowest limit for
its mass turns out to be of $1.66 M_{\odot}$. Other hints for the
existence of compact stars heavier than $2M_{\odot}$ have been
obtained also from the observation and the modeling of short-gamma-ray
bursts (GRB). The SWIFT experiment has detected tens of short-GRB
whose light curves display extended emissions, X-ray flares and
internal plateaus with rapid decay at the end of the plateaus (see
\cite{Lu:2015rta} and Refs. therein).  These observations favor a
model for the inner engine of these events which is based on a rapidly
spinning millisecond magnetar formed from the merger of two neutron
stars. Interestingly, the same stellar objects, but formed after a
supernova, could be the inner engine of long GRBs
\cite{Metzger:2010pp} (see paper 2).  In
\cite{Lasky:2013yaa,Lu:2015rta}, the detailed modeling of the plateaus
seen in short GRBs has provided an important constraint 
on the peak of the expected mass distribution for the post-merger remnant: $M=2.46^{0.13}_{-0.15} M_{\odot}$ (displayed in
Fig. \ref{massa-raggio}). Although this limit is not obtained via a
direct mass measurement and it includes supramassive stars, it represents a strong indication that the
maximum mass of compact stars is significantly larger than
$2M_{\odot}$.

Let us discuss now radii measurements. One has to remark that radii
measurements are much more uncertain than mass measurements and all
the observational constraints are based on specific assumptions made
for modeling the spectra of the X-ray emissions.  In
Refs. \cite{Guillot:2013wu,Guillot:2014lla} the fits on the thermal
emission of 6 quiescent low-mass X-ray binaries, under the assumption
that all of them have the same radius $R$, provide the constraint
$R=9.4\pm1.2$ km. We remark however that these results are under
debate, see Refs. \cite{Lattimer:2013hma,Heinke:2014xaa}. Other
indications of the existence of stars with small radii can be found
from the analysis of X-ray bursts in
Refs.\cite{Ozel:2010fw,Titarchuk:2002im}. In particular, at $1\sigma$,
the analysis of \cite{Ozel:2010fw} indicates radii of about $9.5 \pm
1.5$km and masses of about $M=1.6\pm0.2 M_{\odot}$ (a previous
analysis of 4U1820-30 presented in Ref.\cite{Shaposhnikov:2004zj} has
also found rather small radii: $R=11.2^{+0.4}_{-0.5}$km and
$M=1.29^{+0.19}_{-0.07} M_{\odot}$ ).
For Cyg X-2 a radius of about $9\pm 0.5$km is inferred for the
canonical mass of $1.44\pm 0.06 M_{\odot}$ \cite{Titarchuk:2002im}
while for 4U 1728-34 the suggested ranges are $8.7-9.7$ km for radius
and $1.2-1.6 M_{\odot}$ for the mass \cite{Shaposhnikov:2003cw}.
Also, in the analysis of the X-ray pulsations of SAX J1808.4-3658
one obtains indications of a small radius 
although the uncertainty is still quite large: 
at $3\sigma$, $0.8M_{\odot}<M<1.7M_{\odot}$ and $5 \mathrm{km}<R<13\mathrm{km}$
\cite{Leahy:2007fb,Morsink:2009wv}. 
These data and the mass-radius relation
of hadronic stars (with deltas and hyperons) are displayed in Fig.\ref{massa-raggio2}.

On the other hand, significantly larger radii are obtained by means of
pulse phase-resolved X-ray spectroscopy of PSR J0437-4715
\cite{Bogdanov:2012md}: the radius is constrained to be larger than
$\sim 14$ km at $1\sigma$ confidence level assuming the mass of the star
to be of $1.76 M_{\odot}$ (this value has been obtained via the radio
timing technique in \cite{Verbiest:2008gy}). Similarly, in
Ref. \cite{hambaryan}, a radius larger than about $14$ km is obtained
for the system RX J1856.5−3754 by assuming a mass between $1.5$ and
$1.8 M_{\odot}$.

The existence of very massive compact stars and the possibility that
some neutron stars are very compact represents a serious problem for
the theoretical modeling of the EoS of strongly
interacting matter. 
While massive compact stars imply that the
EoS is stiff, a soft EoS is instead needed
to obtain small radii. This tension is relieved, as proposed in
\cite{Drago:2013fsa}, if one assumes that there are two families of
coexisting compact stars: hadronic stars which can be very compact and
quark stars which can be very massive.  Specifically, the massive PSR
J1614-2230 and PSR J0348+0432, are interpreted in our scenario as
quark stars. Similarly, stars with large radii, as the ones inferred
in the analysis of Refs. \cite{Bogdanov:2012md,hambaryan}, are again
interpreted as quark stars.  On the other hand, the compact stellar objects
such as the ones discusses in the analysis of
\cite{Guillot:2013wu,Guillot:2014lla,Ozel:2010fw,Titarchuk:2002im,Shaposhnikov:2004zj,Shaposhnikov:2003cw} 
would be instead hadronic stars. 
The idea that the equation of state for dense matter 
has a two-phase nature allowing both large and small
compact stars has been suggested also in the analysis of Ref.\cite{Leahy:2008cq}.

In Fig. \ref{massa-raggio} and \ref{massa-raggio2} we display the observational constraints
discussed above and three examples of theoretical 
mass-radius relations (solid and dashed lines) based on the SFHo model with
$\Delta$ only, or with both $\Delta$ and hyperons for the hadronic EoS
(here $x_{\sigma \Delta}=1.15$) and the parametrization presented in
\cite{Fraga:2013qra} for the quark matter EoS. All the
constraints are fulfilled if one assumes that hadronic stars and quark
stars coexist.
Notice that hadronic stars cannot reach masses larger than about
$1.5-1.6 M_{\odot}$ as a result of the softening caused by the
formation of deltas and hyperons. The so called hyperon puzzle, and similarly the delta
puzzle pointed out in \cite{Drago:2014oja}, is easily solved in our
two families scenario: hyperons and deltas do reduce the maximum mass
of compact stars to values significantly smaller than $2M_{\odot}$ but
this fact does not represent a puzzle since the most massive objects
are actually quark stars. Notice that also in the scenarios i) and
ii) discussed in the introduction, massive stars are composed mostly
of quark matter.
 
A natural question concerns the way the quark star branch is
populated.  The stellar configuration for which the solid black line
starts to deviate from the solid green line corresponds to the onset
of hyperons. Once a critical amount of hyperons is present in the
center of the star, nucleation of quark matter can start and can
subsequently trigger the conversion to a quark star. The conversion
occurs because it is energetically convenient: at a fixed value of the
baryonic mass, the gravitational mass of the star on the quark star
branch is smaller than the one on the hadronic star branch (see the
brown line for an example of this conversion process).  The process of
conversion is specifically analyzed in the accompanying paper 2.


\section{Binary systems}
Another useful constraint on the EoS can be obtained
also by studying the double pulsar system J0737-3039
\cite{Burgay:2003jj}. The mass of the so called Pulsar B is of
$1.249\pm0.001 M_{\odot}$. Under the assumption that this pulsar was
formed from an electron capture supernova one can infer a baryonic
mass in the range $1.366-1.375 M_{\odot}$
\cite{Podsiadlowski:2005ig}. In our scenario we can interpret this
star as a hadronic star (which contains $\Delta$ resonances but which
is too light to allow the formation of hyperons). In Fig.\ref{mg-mb},
we display the relation between the gravitational mass and the
baryonic mass for hadronic stars and quark stars.  In the insert, we
show also the constraint of \cite{Podsiadlowski:2005ig}.  Our hadronic
EoS is perfectly in agreement with the analysis of
Ref.\cite{Podsiadlowski:2005ig}: the appearance of delta resonances
gives a small additional contribution to the binding energy of
hadronic stars as compared to nucleonic stars. When considering only
nucleons (see green line), the constraint of
\cite{Podsiadlowski:2005ig} is not fulfilled.  Notice however that
detailed supernova simulations have shown that the uncertainties
associated with the EoS and the wind ablation are such
that the allowed baryonic mass window is shifted towards smaller
values \cite{Kitaura:2005bt}.

\begin{figure}[ptb]
\vskip 1cm
\begin{centering}
\epsfig{file=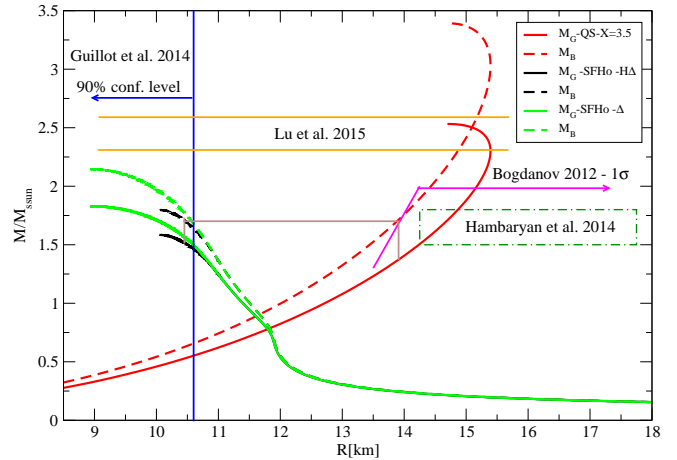,height=8.5cm,width=6cm,angle=-90}
\caption{Gravitational mass-radius (solid lines) and baryonic mass-radius (dashed lines) 
relations for hadronic stars (including only deltas and 
hyperons and deltas ) and quark stars. Some of the most recent observational constraints are also displayed (see text).
The brown lines show that given a hadronic star configuration (in which also hyperons are present), the quark star
with the same baryon mass has a smaller gravitational mass even if its radius is larger. The conversion of the hadronic star
into a quark star would be therefore energetically favored.}
\label{massa-raggio}
\end{centering}
\end{figure}

\begin{figure}[ptb]
\vskip 1cm
\begin{centering}
\epsfig{file=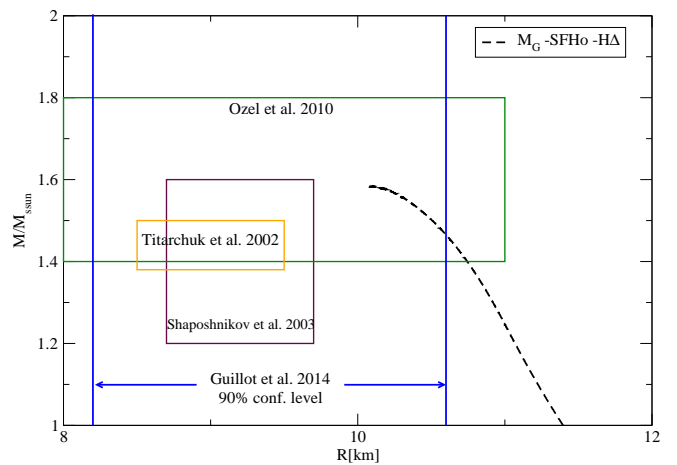,height=8.5cm,width=6cm,angle=-90}
\caption{Mass-radius relation for hadronic stars and observational constraints indicating the existence of very compact stellar objects.}
\label{massa-raggio2}
\end{centering}
\end{figure}

\begin{figure}[ptb]
\vskip 1cm
\begin{centering}
\epsfig{file=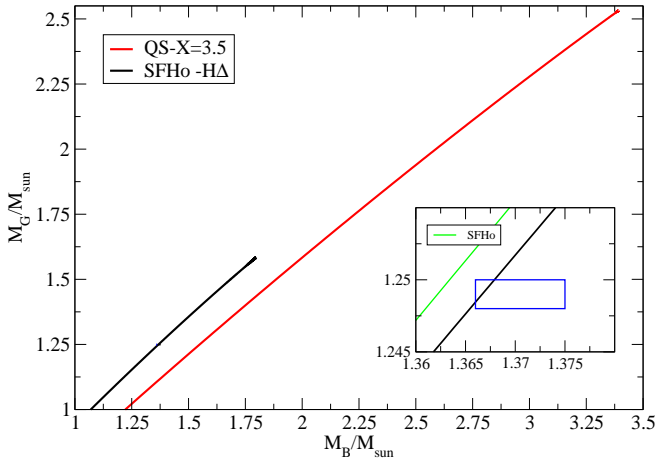,height=8.5cm,width=6cm,angle=-90}
\caption{Relation between gravitational mass and baryonic mass for hadronic stars and quark stars. At fixed baryonic mass
the difference between the gravitational mass of a hadronic star and a quark star is of the order of $0.15 M_{\odot}$. In the insert 
we display also the curve corresponding to nucleonic stars within the SFHo EoS (green line). The slightly larger binding energy obtained
when adding deltas and hyperons allow to fulfill the constraint of \cite{Podsiadlowski:2005ig} (blue box). }
\label{mg-mb}
\end{centering}
\end{figure}

Low-mass X-ray binaries offer another possible hint for the existence
of two families of compact stars. These systems are most probably at
the origin of millisecond pulsars: within the so called recycling
scenario, the neutron star is spun up to milliseconds periods due to
the accretion of mass from its white dwarf companion. Tidal
interactions during the accretion phase are responsible for the
circularization of the orbit and indeed most of the millisecond
pulsars are in circular orbits (with eccentricity $e$ from $10^{-7}$
to $10^{-3}$). However, recently, few examples have been discovered
having a much larger eccentricity such as PSR J2234+06 for which
$e=0.13$ (see other examples in \cite{Knispel:2015qma}). In this
system the white dwarf has a mass of $0.23 M_{\odot}$. The existence of systems with high
eccentricities represents a puzzle in the recycling model of
pulsars. A possible explanation is that the accreting object, at some
point during its evolution, collapses to a more stable configuration
thus increasing abruptly the eccentricity of the binary. In
Ref.~\cite{Freire:2013xma} it has been investigated the scenario of a
rotationally-delayed accretion- induced collapse of a
super-Chandrasekhar mass white dwarf. In Ref.\cite{Jiang:2015gpa}
instead, the accreting star is a neutron star which, due to mass
accretion, converts into a quark star. In our two families scenario
the conversion of a hadronic star to a quark star is necessary once a
sufficient amount of strangeness is formed at the center of the star.
As one can see in Fig.\ref{massa-raggio}, the conversion would occur
for masses of the hadronic star between $\sim 1.35 - 1.6 M_{\odot}$
with an energy released in the conversion given by the difference
between the gravitational mass of the hadronic star $M_H$ and the
gravitational mass of the quark star $M_Q$ computed at the same fixed
baryonic mass: $\Delta M=M_H-M_Q \sim 0.15 M_{\odot}$ (see
Fig. \ref{mg-mb}). The eccentricity is related to the masses of the
hadronic star, of the quark star and of the white dwarf companion
$M_{WD}$ by the following relation: $e=\Delta M/(M_Q+M_{WD})$. It
results that $e\sim 0.1$ if one takes $M_{WD}=0.23 M_{\odot}$,
$M_H=1.55 M_{\odot}$ $M_Q=1.4 M_{\odot}$. A correction to the eccentricity of the order
of $\pm 0.03$ is obtained when considering that during the conversion
the newly born stellar object could get a small kick velocity $v_k$ of
the order of $1$km/sec \cite{Freire:2013xma}. These simple estimates
show that in our model the values of the eccentricities are quite
close to the measured ones. It would be therefore interesting to
investigate more in detail this problem. Future measurements of the masses of the compact stars in those eccentric
systems will be crucial to test our scenario.

\section{Discussion and conclusions}
We have discussed several hints of the existence of two coexisting
families of compact stars, hadronic stars and quark stars. A first
important and widely discussed argument in favor of this scenario is
the necessary appearance of delta resonances and hyperons as the
central density of a hadronic star reaches values larger than about
$2n_0$. The formation of these particles softens the EoS
and reduces the maximum mass with respect to stars made only of
nucleons. How much the maximum mass is reduced due to the appearance
of these particles is the subject of a lively and on-going research
activity in nuclear physics (see 
Refs.\cite{RikovskaStone:2006ta,Oertel:2012qd,Colucci:2013pya,Lopes:2013cpa,Banik:2014qja,vanDalen:2014mqa,Katayama:2014gca,Oertel:2014qza}). 
One one hand, in phenomenological
calculations based on relativistic mean field models the maximum mass
of hyperon stars could still reach the $2 M_{\odot}$ limit (see for
instance \cite{Weissenborn:2011ut,Weissenborn:2011kb,Bednarek:2011gd})
on the other hand in more realistic calculations based on microscopic
nucleon-nucleon interactions, the appearance of hyperons is
accompanied by a strong softening of the EoS which leads
to maximum masses much below $2 M_{\odot}$ and in some case even below
$1.4 M_{\odot}$ \cite{Baldo:1999rq,Vidana:2010ip,Djapo:2008au}.  A
possible way out is that the hadronic EoS is so stiff
that even for the $2 M_{\odot}$ star the central density is below the
threshold for the appearance of hyperons: a possible example has been
given in \cite{Lonardoni:2014bwa} where at a mass $2.09 M_{\odot}$ the
central density is of $3.5 n_0$ and hyperons are not yet formed. This
scenario is realized if the three-body hyperon- nucleon interaction is
sufficiently repulsive. Another possible way to add repulsion between
baryons is the multi-Pomeron exchange potential proposed in
\cite{Yamamoto:2014jga} which, again, would allow the existence of
massive hadronic stars. Unfortunately from the theoretical side one
cannot draw a firm conclusion on the composition of the $2M_{\odot}$
stars.

The crucial quantity to measure is clearly the radius of compact
stars. As already discussed, the presently available analysis are
still affected by large systematic errors. While there are some hints
in favor of the two families scenario (some analysis suggesting the
existence of stars with radii of the order of $10$km and some other
analysis which infer radii of about $14$km) one cannot yet claim that
these measurements have found compelling evidence for the two families
scenario. However, the Neutron star Interior Composition ExploreR
(NICER) instrument, to be launched in 2016, will allow to achieve
$5\%$ precision in neutron-star radius through rotation-resolved
spectroscopy (see for instance \cite{Psaltis:2013fha}). 
Still another 
possibility is based on the future X-ray mission ATHENA+ 
which, combined with accurate distance measurements provided 
by the GAIA experiment, will also allow to obtain information
on masses and radii.  
Another
possible way of extracting the radii is by
analyzing the gravitational-wave emission during the merger of two
compact stars. In particular, the frequency of the dominant
oscillation mode of the post-merger remnant is directly related to the radius of
the non-rotating compact star \cite{Bauswein:2015vxa}.
Again the expected error is of the order of a few hundred meters. 
With such a precision one would possibly finally
establish what the internal composition of compact stars is. Let us
consider the canonical $1.4 M_{\odot}$; one can imagine three possible
outcomes for the measurement of its radius: 
i) radius larger than about $13.5$ km;
ii) radius between $11.5$ and $13.5$ km;
iii) radius smaller than about $11.5$km.
These numbers are also suggested by 
the model-independent study of Ref.\cite{Kurkela:2014vha}. 

In the first case, the large radius implies that the
EoS is stiff. Therefore the central density of the $1.4
M_{\odot}$ star is small and it is possible that even for the most
massive stars the central density is smaller than the threshold 
of formation of hyperons and deltas. Only ``normal''
neutron stars would exist. In the second case, the interpretation
would be much more complicated: several equations of state have been
proposed which can reach the $2M_{\odot}$ limit and predict a radius
of about $12$km for the canonical neutron star mass. Hybrid stars and
hyperonic stars would be both possible (in particular the scenario i)
discussed in the introduction would be favored). 
In the last case,
the only available possibility is the two families scenario that we are
here proposing. Indeed there are no examples of equations of state in
the literature that can at the same time fulfill the $2M_{\odot}$ and
predict such small radii for the canonical neutron star \footnote{In
  Ref. \cite{Guillot:2013wu}, some examples of non-relativistic
  microscopic calculations of nucleonic matter are shown that fulfill
  the $2M_{\odot}$ and, at the same time, predict the existence of very compact
  stars. There are two problems associated with such equations of state and in particular 
  with WFF1\cite{Wiringa:1988tp}: already for the canonical
  mass the central density is of about $4n_0$. At such densities heavy baryons should be included while in WFF1 they are simply neglected.  
  Second: at densities of about
  $1$fm$^{-3}$ it violates causality.}. Within a few years one could
then expect the NICER experiment to give a final answer to the
question about the existence of the two families. 

The scenario of two families of compact stars is based on the
existence of quark stars together with hadronic stars.  In turn the
existence of quark stars implies that the Bodmer-Witten hypothesis is
correct and therefore also small nuggets of quark matter, the so
called strangelets, must exist. We know very little about strangelets:
they are expected to be positively charged and to have a very small
charge to mass ratio, see \cite{Madsen:2004vw}.  Their mass spectrum
is instead completely unknown. As a matter of fact, to date, there is
no experimental evidence of the existence of strangelets.  A possible
way to produce strangelets on earth is through heavy ions experiments:
the search for strangelets at RHIC has produced upper limits of few
$10^{-6}$ to $10^{-7}$ per central Au+Au collision for strangelets
with mass larger than $30$ GeV \cite{Adams:2005cu}.  Notice however
that the production of strangelets in heavy ions collision requires a
net baryon excess \cite{Greiner:1987tg} which is unlikely to be
obtained at RHIC energies. In the context of cosmology a very
interesting hypothesis is that strangelets could represent a type of
``macro dark matter'' which would have been produced by some
post-inflationary process \cite{Jacobs:2014yca}. In this scenario, no
``beyond standard model physics'' would be required to explain the
existence of baryonic dark matter. In astrophysics, strangelets could represent
an important component of cosmic rays. The events that would most
probably produce strangelets are the merger of compact stars with at
least one of the two stars being a quark star, see \cite{Paulucci:2014vna} for a discussion on the fragmentation of quark matter into strangelets. 
The search for strangelets in the lunar soil has provided no evidence of them from
A=42 to A=70 and for nuclear charges of 5, 6, 8, 9 and 11
\cite{Han:2009sj}. New limits on the mass and the flux of cosmic
strangelets are expected to be available in the near future thanks to
the AMS-02 experiment.

From the theoretical side, to date, there has been only one detailed
simulation of the merger of two quark stars \cite{Bauswein:2008gx}
while the possibility of neutron star - quark star and black hole -
quark star merger have not yet been considered.  An unexpected result
of these simulations is that in many cases, after the merger, a prompt
collapse to a black hole occurs and basically no quark matter is
ejected. In particular, this occurs for values of the total mass of
the merger larger than about $2.5-3 M_{\odot}$. In our scenario
quark stars have masses larger than about $1.35 M_{\odot}$
and it is rather difficult to avoid a prompt collapse. 
It is therefore possible
that even if quark matter is absolutely stable the flux of strangelets
is vanishingly small and not all compact stars convert into quark
stars as it would result if the cosmic strangelets pollution would be
significant \cite{Friedman:1990qz}. The two families of compact stars could indeed coexist.

\vskip 0.5cm
 
A.D. would like to thank
V. Hislop for the moral support during the preparation of the paper. 
G.P. acknowledges financial support from the Italian Ministry of Research through the program 
 Rita Levi Montalcini .

%

\begin{thebibliography}{88}

\bibitem{Adams:2005dq}
J.~Adams et~al. (STAR), Nucl. Phys. \textbf{A757}, 102 (2005),
  \texttt{nucl-ex/0501009}

\bibitem{Demorest:2010bx}
P.~Demorest, T.~Pennucci, S.~Ransom, M.~Roberts, J.~Hessels, Nature
  \textbf{467}, 1081 (2010)

\bibitem{Antoniadis:2013pzd}
J.~Antoniadis, P.C. Freire, N.~Wex, T.M. Tauris, R.S. Lynch et~al., Science
  \textbf{340}, 6131 (2013)

\bibitem{Chen:2011my}
H.~Chen, M.~Baldo, G.F. Burgio, H.J. Schulze, Phys. Rev. \textbf{D84}, 105023
  (2011), \texttt{1107.2497}

\bibitem{Bonanno:2011ch}
L.~Bonanno, A.~Sedrakian, Astron. Astrophys. \textbf{539}, A16 (2012),
  \texttt{1108.0559}

\bibitem{Zdunik:2012dj}
J.L. Zdunik, P.~Haensel, Astron. Astrophys. \textbf{551}, A61 (2013),
  \texttt{1211.1231}

\bibitem{Kurkela:2014vha}
A.~Kurkela, E.S. Fraga, J.~Schaffner-Bielich, A.~Vuorinen, Astrophys. J.
  \textbf{789}, 127 (2014), \texttt{1402.6618}

\bibitem{Benic:2014jia}
S.~Benic, D.~Blaschke, D.E. Alvarez-Castillo, T.~Fischer, S.~Typel, Astron.
  Astrophys. \textbf{577}, A40 (2015), \texttt{1411.2856}

\bibitem{Drago:2013fsa}
A.~Drago, A.~Lavagno, G.~Pagliara, Phys.Rev. \textbf{D89}, 043014 (2014),
  \texttt{1309.7263}

\bibitem{Glendenning:1991es}
N.~Glendenning, S.~Moszkowski, Phys.Rev.Lett. \textbf{67}, 2414 (1991)

\bibitem{Lattimer:2012xj}
J.M. Lattimer, Y.~Lim, ApJ. 771, \textbf{51} (2013), \texttt{1203.4286}
\bibitem{Tsang:2012se}
  M.~B.~Tsang {\it et al.},
  Phys.\ Rev.\ C {\bf 86} (2012) 015803.


\bibitem{Steiner:2004fi}
A.W. Steiner, M.~Prakash, J.M. Lattimer, P.J. Ellis, Phys.Rept. \textbf{411},
  325 (2005), \texttt{nucl-th/0410066}

\bibitem{Steiner:2012rk}
A.W. Steiner, M.~Hempel, T.~Fischer, Astrophys.J. \textbf{774}, 17 (2013),
  \texttt{1207.2184}

\bibitem{Raduta:2014lja}
A.R. Raduta, F.~Gulminelli, M.~Oertel (2014), \texttt{1406.0395}

\bibitem{Schaffner:1993nn}
J.~Schaffner, C.B. Dover, A.~Gal, C.~Greiner, H.~Stoecker, Phys.Rev.Lett.
  \textbf{71}, 1328 (1993)

\bibitem{Schaffner:1995th}
J.~Schaffner, I.N. Mishustin, Phys.Rev. \textbf{C53}, 1416 (1996)

\bibitem{Zabrodin:2009fz}
E.E. Zabrodin, I.C. Arsene, J.~Bleibel, M.~Bleicher, L.V. Bravina et~al.,
  J.Phys. \textbf{G36}, 064065 (2009)

\bibitem{Hofmann:1994gn}
M.~Hofmann, R.~Mattiello, H.~Sorge, H.~Stoecker, W.~Greiner, Phys.Rev.
  \textbf{C51}, 2095 (1995)

\bibitem{Bass:1998vz}
S.~Bass, M.~Gyulassy, H.~Stoecker, W.~Greiner, J.Phys. \textbf{G25}, R1 (1999)

\bibitem{Lavagno:2010ah}
A.~Lavagno, Phys.Rev. \textbf{C81}, 044909 (2010)

\bibitem{Lavagno:2012bn}
A.~Lavagno, D.~Pigato, Phys.Rev. \textbf{C86}, 024917 (2012)

\bibitem{Huber:1997mg}
H.~Huber, F.~Weber, M.~Weigel, C.~Schaab, Int.J.Mod.Phys. \textbf{E7}, 301
  (1998), \texttt{nucl-th/9711025}

\bibitem{Xiang:2003qz}
H.~Xiang, G.~Hua, Phys.Rev. \textbf{C67}, 038801 (2003)

\bibitem{Chen:2007kxa}
Y.~Chen, H.~Guo, Y.~Liu, Phys.Rev. \textbf{C75}, 035806 (2007)

\bibitem{Chen:2009am}
Y.~Chen, Y.~Yuan, Y.~Liu, Phys.Rev. \textbf{C79}, 055802 (2009)

\bibitem{Schurhoff:2010ph}
T.~Schurhoff, S.~Schramm, V.~Dexheimer, Astrophys.J. \textbf{724}, L74 (2010)

\bibitem{Glendenning:1984jr}
N.~Glendenning, Astrophys.J. \textbf{293}, 470 (1985)

\bibitem{Kosov:1998gp}
D.~Kosov, C.~Fuchs, B.~Martemyanov, A.~Faessler, Phys.Lett. \textbf{B421}, 37
  (1998)

\bibitem{Jin:1994vw}
X.m. Jin, Phys.Rev. \textbf{C51}, 2260 (1995)

\bibitem{Oset:1987re}
E.~Oset, L.~Salcedo, Nucl.Phys. \textbf{A468}, 631 (1987)

\bibitem{Koch:1985qz}
J.~Koch, N.~Ohtsuka, Nucl.Phys. \textbf{A435}, 765 (1985)

\bibitem{Wehrberger:1989cd}
K.~Wehrberger, C.~Bedau, F.~Beck, Nucl.Phys. \textbf{A504}, 797 (1989)

\bibitem{O'Connell:1990zg}
J.~O'Connell, R.~Sealock, Phys.Rev. \textbf{C42}, 2290 (1990)

\bibitem{Alberico:1994sx}
W.~Alberico, G.~Gervino, A.~Lavagno, Phys.Lett. \textbf{B321}, 177 (1994)

\bibitem{Horikawa:1980cv}
Y.~Horikawa, M.~Thies, F.~Lenz, Nucl.Phys. \textbf{A345}, 386 (1980)

\bibitem{Nakamura:2009iq}
S.~Nakamura, T.~Sato, T.S. Lee, B.~Szczerbinska, K.~Kubodera, Phys.Rev.
  \textbf{C81}, 035502 (2010), \texttt{0910.1057}

\bibitem{Drago:2014oja}
A.~Drago, A.~Lavagno, G.~Pagliara, D.~Pigato, Phys. Rev. \textbf{C90}, 065809
  (2014), \texttt{1407.2843}

\bibitem{PhysRevC.92.015802}
B.J. Cai, F.J. Fattoyev, B.A. Li, W.G. Newton, Phys. Rev. C \textbf{92}, 015802
  (2015)

\bibitem{Li:2015hfa}
B.A. Li (2015), \texttt{1507.03279}

\bibitem{Benlliure:2015qea}
J.~Benlliure et~al., JPS Conf. Proc. \textbf{6}, 020039 (2015)

\bibitem{Li:1997yh}
Z.X. Li, G.J. Mao, Y.Z. Zhuo, W.~Greiner, Phys.Rev. \textbf{C56}, 1570 (1997)

\bibitem{Typel:2009sy}
S.~Typel, G.~Ropke, T.~Klahn, D.~Blaschke, H.~Wolter, Phys.Rev. \textbf{C81},
  015803 (2010), \texttt{0908.2344}

\bibitem{Cai:2015hya}
B.J. Cai, F.J. Fattoyev, B.A. Li, W.G. Newton, Phys. Rev. \textbf{C92}, 015802
  (2015), \texttt{1501.01680}

\bibitem{Prakash:1996xs}
M.~Prakash, I.~Bombaci, M.~Prakash, P.J. Ellis, J.M. Lattimer, R.~Knorren,
  Phys. Rept. \textbf{280}, 1 (1997), \texttt{nucl-th/9603042}

\bibitem{Buballa:2003qv}
M.~Buballa, Phys. Rept. \textbf{407}, 205 (2005), \texttt{hep-ph/0402234}

\bibitem{Kurkela:2009gj}
A.~Kurkela, P.~Romatschke, A.~Vuorinen, Phys.Rev. \textbf{D81}, 105021 (2010),
  \texttt{0912.1856}

\bibitem{Fraga:2013qra}
E.S. Fraga, A.~Kurkela, A.~Vuorinen, Astrophys.J. \textbf{781}, L25 (2014),
  \texttt{1311.5154}

\bibitem{Weissenborn:2011qu}
  S.~Weissenborn, I.~Sagert, G.~Pagliara, M.~Hempel and J.~Schaffner-Bielich,
  Astrophys.\ J.\  {\bf 740} (2011) L14.


\bibitem{Zacchi:2015lwa}
  A.~Zacchi, R.~Stiele and J.~Schaffner-Bielich,
  Phys.\ Rev.\ D {\bf 92} (2015) 4,  045022.



\bibitem{vanKerkwijk:2010mt}
M.~van Kerkwijk, R.~Breton, S.~Kulkarni, Astrophys.J. \textbf{728}, 95 (2011)

\bibitem{Lu:2015rta}
H.J. Lü, B.~Zhang, W.H. Lei, Y.~Li, P.D. Lasky, Astrophys.J. \textbf{805}, 89
  (2015), \texttt{1501.02589}

\bibitem{Metzger:2010pp}
B.~Metzger, D.~Giannios, T.~Thompson, N.~Bucciantini, E.~Quataert (2010)

\bibitem{Lasky:2013yaa}
P.D. Lasky, B.~Haskell, V.~Ravi, E.J. Howell, D.M. Coward, Phys.Rev.
  \textbf{D89}, 047302 (2014), \texttt{1311.1352}

\bibitem{Guillot:2013wu}
S.~Guillot, M.~Servillat, N.A. Webb, R.E. Rutledge (2013), \texttt{1302.0023}

\bibitem{Guillot:2014lla}
S.~Guillot, R.E. Rutledge, Astrophys.J. \textbf{796}, L3 (2014),
  \texttt{1409.4306}

\bibitem{Lattimer:2013hma}
J.M. Lattimer, A.W. Steiner (2013), \texttt{1305.3242}

\bibitem{Heinke:2014xaa}
C.O. Heinke et~al., Mon. Not. Roy. Astron. Soc. \textbf{444}, 443 (2014),
  \texttt{1406.1497}

\bibitem{Ozel:2010fw}
F.~Ozel, G.~Baym, T.~Guver, Phys.Rev. \textbf{D82}, 101301 (2010),
  \texttt{1002.3153}

\bibitem{Titarchuk:2002im}
L.~Titarchuk, N.~Shaposhnikov, Astrophys. J. \textbf{570}, L25 (2002),
  \texttt{astro-ph/0203432}

\bibitem{Shaposhnikov:2004zj}
N.~Shaposhnikov, L.~Titarchuk, Astrophys. J. \textbf{606}, L57 (2004),
  \texttt{astro-ph/0403488}

\bibitem{Shaposhnikov:2003cw}
N.~Shaposhnikov, L.~Titarchuk, F.~Haberl, Astrophys. J. \textbf{593}, L35
  (2003), \texttt{astro-ph/0307215}

\bibitem{Leahy:2007fb}
D.A. Leahy, S.M. Morsink, C.~Cadeau, Astrophys. J. \textbf{672}, 1119 (2008),
  \texttt{astro-ph/0703287}

\bibitem{Morsink:2009wv}
  S.~M.~Morsink and D.~A.~Leahy,
  Astrophys.\ J.\  {\bf 726} (2011) 56
  [arXiv:0911.0887 [astro-ph.HE]].

\bibitem{Bogdanov:2012md}
S.~Bogdanov, Astrophys.J. \textbf{762}, 96 (2013), \texttt{1211.6113}

\bibitem{Verbiest:2008gy}
J.P.W. Verbiest, M.~Bailes, W.~van Straten, G.B. Hobbs, R.T. Edwards, R.N.
  Manchester, N.D.R. Bhat, J.M. Sarkissian, B.A. Jacoby, S.R. Kulkarni,
  Astrophys. J. \textbf{679}, 675 (2008), \texttt{0801.2589}

\bibitem{hambaryan}
V.~Hambaryan, R.~Neuhaeuser, V.~Suleimanov, K.~Werner, Journal of Physics:
  Conf.Ser. \textbf{496}, 012015 (2014)

\bibitem{Leahy:2008cq}
  D.~A.~Leahy, S.~M.~Morsink, Y.~Y.~Chung and Y.~Chou,
  Astrophys.\ J.\  {\bf 691} (2009) 1235
  [arXiv:0806.0824 [astro-ph]].

\bibitem{Burgay:2003jj}
M.~Burgay et~al., Nature \textbf{426}, 531 (2003), \texttt{astro-ph/0312071}

\bibitem{Podsiadlowski:2005ig}
P.~Podsiadlowski, J.D.M. Dewi, P.~Lesaffre, J.C. Miller, W.G. Newton, J.R.
  Stone, Mon. Not. Roy. Astron. Soc. \textbf{361}, 1243 (2005),
  \texttt{astro-ph/0506566}

\bibitem{Kitaura:2005bt}
F.S. Kitaura, H.T. Janka, W.~Hillebrandt, Astron. Astrophys. \textbf{450}, 345
  (2006), \texttt{astro-ph/0512065}

\bibitem{Knispel:2015qma}
B.~Knispel et~al., Astrophys. J. \textbf{806}, 140 (2015), \texttt{1504.03684}

\bibitem{Freire:2013xma}
P.C.C. Freire, T.M. Tauris, Mon.Not.Roy.Astron.Soc. \textbf{438}, 86 (2014),
  \texttt{1311.3478}

\bibitem{Jiang:2015gpa}
L.~Jiang, X.D. Li, J.~Dey, M.~Dey, Astrophys.J. \textbf{807}, 41 (2015),
  \texttt{1505.04644}

\bibitem{RikovskaStone:2006ta}
  J.~Rikovska-Stone, P.~A.~M.~Guichon, H.~H.~Matevosyan and A.~W.~Thomas,
  Nucl.\ Phys.\ A {\bf 792} (2007) 341.

\bibitem{Oertel:2012qd}
  M.~Oertel, A.~F.~Fantina and J.~Novak,
  Phys.\ Rev.\ C {\bf 85} (2012) 055806.

\bibitem{Colucci:2013pya}
  G.~Colucci and A.~Sedrakian,
  Phys.\ Rev.\ C {\bf 87} (2013) 055806.

\bibitem{Lopes:2013cpa}
  L.~L.~Lopes and D.~P.~Menezes,
  Phys.\ Rev.\ C {\bf 89} (2014) 2,  025805.

\bibitem{Banik:2014qja}
  S.~Banik, M.~Hempel and D.~Bandyopadhyay,
  Astrophys.\ J.\ Suppl.\  {\bf 214} (2014) 2,  22.

\bibitem{vanDalen:2014mqa}
  E.~N.~E.~van Dalen, G.~Colucci and A.~Sedrakian,
  Phys.\ Lett.\ B {\bf 734} (2014) 383.

\bibitem{Katayama:2014gca}
  T.~Katayama and K.~Saito,
  arXiv:1410.7166 [nucl-th].

\bibitem{Oertel:2014qza}
  M.~Oertel, C.~Providência, F.~Gulminelli and A.~R.~Raduta,
  J.\ Phys.\ G {\bf 42} (2015) 7,  075202.


\bibitem{Weissenborn:2011ut}
S.~Weissenborn, D.~Chatterjee, J.~Schaffner-Bielich, Phys.Rev. \textbf{C85},
  065802 (2012)

\bibitem{Weissenborn:2011kb}
S.~Weissenborn, D.~Chatterjee, J.~Schaffner-Bielich, Nucl.Phys. \textbf{A881},
  62 (2012), \texttt{1111.6049}

\bibitem{Bednarek:2011gd}
I.~Bednarek, P.~Haensel, J.~Zdunik, M.~Bejger, R.~Manka (2011),
  \texttt{1111.6942}

\bibitem{Baldo:1999rq}
M.~Baldo, G.~Burgio, H.~Schulze, Phys.Rev. \textbf{C61}, 055801 (2000),
  \texttt{nucl-th/9912066}

\bibitem{Vidana:2010ip}
I.~Vidana, D.~Logoteta, C.~Providencia, A.~Polls, I.~Bombaci, Europhys.Lett.
  \textbf{94}, 11002 (2011), \texttt{1006.5660}

\bibitem{Djapo:2008au}
H.~Djapo, B.J. Schaefer, J.~Wambach, Phys. Rev. \textbf{C81}, 035803 (2010),
  \texttt{0811.2939}

\bibitem{Lonardoni:2014bwa}
D.~Lonardoni, A.~Lovato, S.~Gandolfi, F.~Pederiva, Phys. Rev. Lett.
  \textbf{114}, 092301 (2015).

\bibitem{Yamamoto:2014jga}
Y.~Yamamoto, T.~Furumoto, N.~Yasutake, T.A. Rijken, Phys. Rev. \textbf{C90},
  045805 (2014).

\bibitem{Psaltis:2013fha}
D.~Psaltis, F.~Özel, D.~Chakrabarty, Astrophys. J. \textbf{787}, 136 (2014).

\bibitem{Bauswein:2015vxa}
A.~Bauswein, N.~Stergioulas, H.T. Janka (2015), \texttt{1508.05493}

\bibitem{Wiringa:1988tp}
R.B. Wiringa, V.~Fiks, A.~Fabrocini, Phys.Rev. \textbf{C38}, 1010 (1988).

\bibitem{Madsen:2004vw}
J.~Madsen, Phys.Rev. \textbf{D71}, 014026 (2005).

\bibitem{Adams:2005cu}
B.I. Abelev et~al. (STAR), Phys. Rev. \textbf{C76}, 011901 (2007),
  \texttt{nucl-ex/0511047}

\bibitem{Greiner:1987tg}
C.~Greiner, P.~Koch, H.~Stoecker, Phys. Rev. Lett. \textbf{58}, 1825 (1987)

\bibitem{Jacobs:2014yca}
D.M. Jacobs, G.D. Starkman, B.W. Lynn, Mon. Not. Roy. Astron. Soc.  (2014),
  \texttt{1410.2236}

\bibitem{Paulucci:2014vna}
L.~Paulucci, J.E. Horvath, Phys. Lett. \textbf{B733}, 164 (2014),
  \texttt{1405.1777}

\bibitem{Han:2009sj}
K.~Han, J.~Ashenfelter, A.~Chikanian, W.~Emmet, L.E. Finch, A.~Heinz,
  J.~Madsen, R.D. Majka, B.~Monreal, J.~Sandweiss, Phys. Rev. Lett.
  \textbf{103}, 092302 (2009), \texttt{0903.5055}

\bibitem{Bauswein:2008gx}
A.~Bauswein, H.T. Janka, R.~Oechslin, G.~Pagliara, I.~Sagert et~al.,
  Phys.Rev.Lett. \textbf{103}, 011101 (2009)

\bibitem{Friedman:1990qz}
J.L. Friedman, R.R. Caldwell, Phys. Lett. \textbf{B264}, 143 (1991)

\end{thebibliography}
%
%
%

\end{document}